\documentclass[a4paper,11pt]{article}
\usepackage{amssymb,amsmath,amsthm,amsfonts}
\usepackage{bookmark}
\usepackage{mathrsfs}
\usepackage{multirow}
\usepackage{graphicx}
\usepackage{authblk}
\usepackage{indentfirst}
\usepackage{multicol}
\usepackage{tabu}
\usepackage{tabularx}
\usepackage{url}
\usepackage{fancyhdr}
\usepackage[numbers]{natbib}
\usepackage[all]{xy}
\usepackage{mathtools}
\usepackage[english]{babel}
\usepackage{colortbl}
\usepackage{caption}
\usepackage{hyperref}
\usepackage{setspace}
\usepackage[mathlines]{lineno}

\usepackage{xr}
\makeatletter
    \newcommand*{\addFileDependency}[1]{
    \typeout{(#1)}
    \@addtofilelist{#1}
    \IfFileExists{#1}{}{\typeout{No file #1.}}
    }
\makeatother

\title{Rapid response to fast viral evolution using AlphaFold 3-assisted topological deep learning }

\author{JunJie Wee$^{1}$, and Guo-Wei Wei$^{1,2,3,\ast}$\\
\normalsize{$^{1}$Department of Mathematics, Michigan State University, East Lansing, MI 48824, USA}\\
\normalsize{$^{2}$Department of Biochemistry and Molecular Biology,}\\ 
\normalsize{Michigan State University, East Lansing, MI 48824, USA}\\
\normalsize{$^{3}$Department of Electrical and Computer Engineering,} \\
\normalsize{Michigan State University, East Lansing, MI 48824, USA}\\
\normalsize{$^\ast$ Address correspondences to Guo-Wei Wei. E-mail: weig@msu.edu}
}

\date{}

\begin{document}




\maketitle

\begin{abstract}
The fast evolution of SARS-CoV-2 and other infectious viruses poses a grand challenge  to the rapid response in terms of viral tracking, diagnostics,  and design and manufacture of monoclonal antibodies (mAbs) and vaccines,  which are both time-consuming and costly. This underscores the need for efficient computational approaches. Recent advancements, like topological deep learning (TDL), have introduced powerful tools for forecasting emerging dominant variants, yet they require deep mutational scanning (DMS) of viral surface proteins and associated three-dimensional (3D) protein-protein interaction (PPI) complex structures.  We propose an AlphaFold 3 (AF3)-assisted multi-task topological Laplacian (MT-TopLap) strategy to address this need. MT-TopLap combines deep learning with topological data analysis (TDA) models, such as persistent Laplacians (PL) to extract detailed topological and geometric characteristics of PPIs, thereby enhancing the prediction of DMS and binding free energy (BFE) changes upon virus mutations. Validation with four experimental DMS datasets of  SARS-CoV-2   spike receptor-binding domain (RBD) and the human angiotensin-converting enzyme-2 (ACE2) complexes  
indicates that our AF3 assisted MT-TopLap strategy maintains robust performance, with only an average 1.1\% decrease in Pearson correlation coefficients (PCC) and an average 9.3\% increase in root mean square errors (RMSE), compared with the use of experimental structures. Additionally, AF3-assisted MT-TopLap achieved a PCC of 0.81 when tested with a SARS-CoV-2 HK.3 variant DMS dataset, confirming its capability to accurately predict BFE changes and adapt to new experimental data, thereby showcasing its potential for  rapid and effective response to fast viral evolution.

\end{abstract}

\newpage
\begin{spacing}{0.1}
\tableofcontents
\end{spacing}

\newpage

\section{Introduction}

According to the Centers for Disease Control and Prevention (CDC)\cite{flu-vaccines-work2024}, the effectiveness of flu vaccines has been much lower than 50\% over the past decade, mostly due to fast evolution of flu viruses and the relatively slow process of vaccine design and manufacture. It takes about six months for pharmaceutical makers to produce a flu vaccine, whereas, during this period, the flu virus typically has undergone some crucial mutations at its surface protein. As such, it is extremely valuable to predict the emerging dominant viral variants so that new vaccines can be designed and put into manufacture before a viral variant becomes dominant. During the COVID-19 pandemic, the emerging dominance of Omicron BA.2 \cite{chen2022omicron} and BA.4/BA.5 \cite{chen2022persistent} was accurately forecast nearly two months in advance of the World Health Organization (WHO) announcement, offering a life-saving early warning to society. These predictions were based on two natural selection mechanisms of SARS-Cov-2 evolution, namely infectivity strengthening \cite{chen2020mutations} and antibody resistance (or vaccine breakthrough) \cite{wang2021mechanisms}. Technically, the accurate forecasting of emerging dominant variants was achieved via the integration of artificial intelligence (AI), topological deep learning (TDL), genotyping of viral genomes extracted from patients,  the deep mutational scanning (DMS) of SARS-CoV-2 receptor binding domain (RBD) in complex with human angiotensin-converting enzyme 2 (ACE2), and the three-dimensional (3D) RBD-ACE2 complex structure. Among these, the DMS experiments are typically a bottleneck in  response to fast evolving viruses \cite{starr2022deep,starr2020deep,dadonaite2024spike}. Indeed, the earliest experimental DMS on the SARS-CoV-2 spike protein RBD mutations was not available until September 2020, nine months after the outbreak of the COVID-19 pandemic \cite{starr2020deep}. Additionally, during the global spread of infectious viruses \cite{bergquist2020covid}, rapid viral evolution poses significant challenges for experimental DMS to keep pace.

Since experiments are time-consuming and high-cost, much effort has been given to the  development of effective computational methods for predicting mutational impacts on viral surface protein and human receptor interactions or protein-protein interactions (PPIs) \cite{geng2019finding,pandurangan2020prediction,geng2019isee,li2021saambe,siebenmorgen2020computational,dehouck2013beatmusic,romero2022ppi,xue2016prodigy,yang2023area,wang2020topology}. Remarkably, the TDL approach predicted  in-silico DMS of SARS-CoV-2 spike RBD in May 2020 \cite{chen2020mutations}, four months earlier than the first experimental DMS \cite{starr2020deep}. This study identified critical mutation sites at spike protein residue positions 452 and 501, which were predicted to result in significantly more infectious SARS-CoV-2 variants \cite{chen2020mutations}. These sites were subsequently confirmed as key hotspots in several prevailing SARS-CoV-2 variants, including Alpha, Gamma, Delta, Beta, Theta, Mu, Omicron, BA.2, BA.4, BA.5, and all later variants. Nonetheless, experimental DMS is needed for highly accurate and reliable AI-based predictions, such as the aforementioned forecasting of Omicron BA.2 and BA.4/BA.5's emerging dominance \cite{chen2022omicron,chen2022persistent}.

Introduced in 2017, TDL combines topological data analysis (TDA) with deep learning techniques \cite{cang2017topologynet}. A significant factor in this success is persistent homology \cite{edelsbrunner2008persistent,zomorodian2004computing}, which was used to generate topological fingerprints for predicting binding free energy (BFE) changes upon mutation in  PPIs complexes \cite{wang2020topology}. Notably, one of the earliest persistent homology models to integrate with deep mutational scanning is TopNetmAb, which offered accurate predictions of BFE changes upon mutations in the RBD and ACE2 complexes \cite{chen2020mutations,chen2021prediction}.

However, persistent homology has limitations, such as the lack of description of non-topological shape evolution in data. Recently, persistent Laplacian (PL) has been introduced to address these limitations of existing TDA methods \cite{wang2020persistent}. It extracts both topological and geometric information to understand the shape and structure of complex and high-dimensional data. PL is deeply rooted in spectral theory and is part of the family of persistent topological Laplacians, like 
persistent sheaf Laplacians \cite{wei2024persistent}  and persistent Mayer Laplacians \cite{shen2024persistent}. The harmonic spectra of PL can fully recover the Betti numbers of persistent homology \cite{edelsbrunner2008persistent,zomorodian2004computing}, a crucial tool in the early days of TDA.

PL has proven effective in scenarios where standard approaches struggle, and it enhances situations where they perform well by providing harmonic and non-harmonic spectral features for TDL models \cite{chen2022persistent, wee2022persistent,liu2022hom}. PL is also used in the latest integration between TDL and pre-trained evolutionary scale modeling (ESM) transformer features \cite{rives2021biological} to predict mutation-induced protein solubility changes, establishing a state-of-the-art method for predicting protein-protein binding free energy (BFE) changes \cite{wee2024integration}.

Apart from experimental DMS data, TDL or TopNetmAb also employs the SKEMPI 2.0 database \cite{jankauskaite2019skempi}, an experimental database for PPI BFE changes upon mutation. SKEMPI 2.0 is used as training data with a multi-task deep learning approach that simultaneously learns single-site mutational patterns with experimental BFE changes and DMS profiles with experimental enrichment ratios. Our recent study on multi-task topological Laplacian (MT-TopLap) \cite{wee2024preventing}, a multi-task deep learning model built by integrating persistent Laplacian, auxiliary, and pre-trained ESM-2 transformer features, revealed a 5\% and 21\% improvement in root mean square error (RMSE) compared to earlier topology-based models like TopLapNet \cite{chen2022persistent} and non-topology-based models like mCSM-PPI2 \cite{rodrigues2019mcsm} when predicting BFE changes upon mutation.

Finally, one important factor in ensuring the success of TDL or MT-TopLap for DMS predictions is the availability of high  quality 3D structures of PPI complexes. MT-TopLap requires the  3D  structures of new viral variants to achieve  accurate predictions. Such structures are typically not available until many months later during pandemics. Without these structures, new experimental DMS data cannot be utilized in TDL models. The rapid mutation rate of SARS-CoV-2 exacerbates this issue, leading to a shortage of both DMS and 3D structural data for emerging SARS-CoV-2 variants. This lack of 3D structures directly poses a significant challenge for TDL and many other machine learning methods, hindering the ability of in-silico DMS to keep pace with viral evolution.

The challenge of lacking 3D structures may be addressed by AlphaFold 3 (AF3), which represents a groundbreaking advancement in the field of computational biology, particularly in the prediction of antibody-antigen or protein-protein interactions (PPIs). Developed by Google DeepMind and Isomorphic Labs \cite{abramson2024accurate}, AF3 builds upon the remarkable success of its predecessors by introducing significant enhancements in protein structure prediction. One of the most notable features of AF3 is its ability to accurately predict the structures of PPI complexes, including those involving DNA, RNA, antibodies, and antigens. Accurate prediction of PPIs is essential for understanding the molecular mechanisms underlying various diseases, developing targeted therapies, and advancing protein engineering efforts.

The capabilities of AF3 in predicting PPIs mark a substantial leap from previous iterations, which primarily focused on predicting the 3D structures of individual proteins. In 2019, AlphaFold’s \cite{jumper2021highly} exceptional performance in predicting 58.1\% of the test protein structures generated excitement among researchers about the future of AI-driven protein structure prediction. DeepFragLib by Wang et al. \cite{wang2019improved} then made a significant advancement in ab initio protein structure prediction. The introduction of AlphaFold marked a transformative shift in how we model protein structures and their interactions \cite{jumper2021highly}. AlphaFold has since unlocked numerous possibilities in protein folding, protein engineering, and design \cite{wei2019protein,qiu2023persistent,lin2023evolutionary,mirdita2022colabfold,tunyasuvunakool2021highly,dejnirattisai2022sars,nunes2023alphafold2}. In October 2024, the Nobel Prize in Chemistry was awarded to David Baker, Demis Hassabis, and John Jumper for their revolutionary contributions to AlphaFold.


In this work, we introduce AF3-assisted MT-TopLap to accurately predict the BFE changes upon DMS mutations  of SARS-CoV-2 RBD-ACE2 variants. Specifically, we leverage AF3's predicted structures to extract persistent Laplacian features. As such, we can speed up in-silico DMS predictions by bypassing the experimental determination of PPI complex structures. As a result, the proposed AF3-assisted MT-TopLap can serve the need for a rapid response to fast viral evolution.

We validate our AF3-assisted MT-TopLap approach with four experimental RBD-ACE2 DMS datasets of SARS-CoV-2 and its variants and show that AF3-assisted MT-TopLap maintains strong performance in predicting BFE changes upon DMS mutations, with an average decrease of only 1.1\% in Pearson correlation coefficient (PCC) and an average increase of only 9.3\% in RMSE, compared to the use of experimental RBD-ACE2 structures. Furthermore, we perform fine-tuning validation of AF3-assisted MT-TopLap using an experimental RBD-ACE2 DMS dataset of SARS-CoV-2 HK.3   \cite{taylor2024deep}. AF3-assisted MT-TopLap achieves a PCC of 0.81 and RMSE of 1.10 for this fine-tuning validation. Using AF3's SARS-CoV-2 RBD-ACE2 complexes  extends MT-TopLap's effectiveness in predicting viral mutation impacts, tracking virus evolution, forecasting emerging dominant variants,  and guiding the development of new vaccines.

\section{Results}

In this section, we assess the performance of AF3-assisted MT-TopLap in predicting four SARS-CoV-2 RBD-ACE2 DMS datasets. 
 MT-TopLap integrates persistent Laplacian, auxiliary, and pre-trained transformer features to predict BFE changes upon RBD DMS mutations. 
Using alternative AF3's 3D RBD-ACE2 complex structures, we demonstrate that AF3-assisted MT-TopLap can effectively predict the BFE changes upon RBD DMS mutations for SARS-CoV-2 evolution. 
Specifically, we first collected four experimental RBD-ACE2 DMS datasets, each paired with an experimental 3D SARS-CoV-2 RBD-ACE2 complex structure. Meanwhile, we also create alternative AF3 3D RBD-ACE2 complex structures from the AlphaFold 3 Server. We evaluate how these alternative AF3 3D RBD-ACE2 complex structures impact our MT-TopLap model's predictions of BFE changes upon the DMS of RBD, which involves the systematic mutations of each RBD residue into 19 other residues, resulting in thousands of mutations.

\begin{figure}[!htpb]
	\centering
	\includegraphics[width=\textwidth]{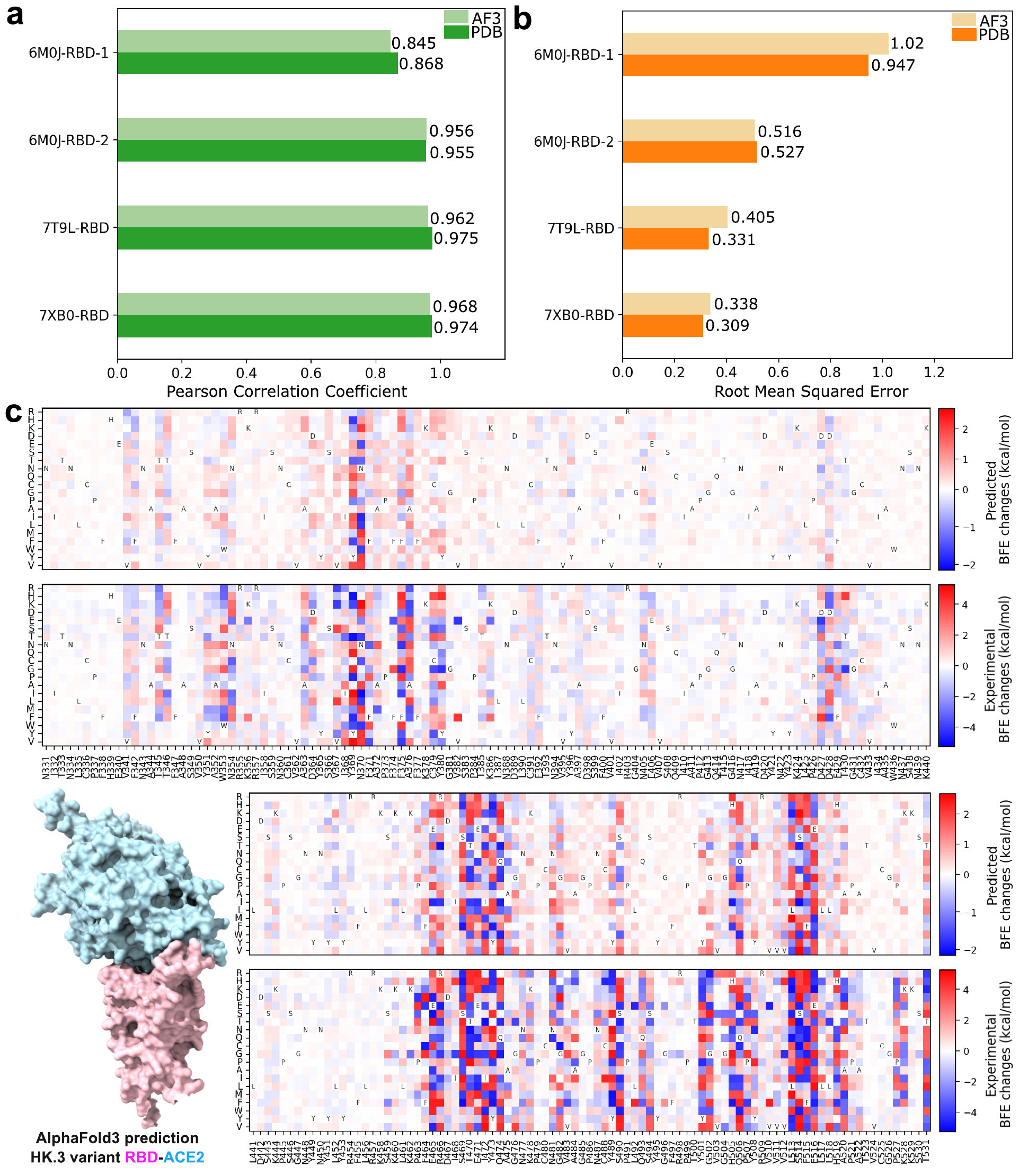}
	\caption{(a): Comparison of AF3-assisted MT-TopLap's   Pearson correlation coefficient (PCC) values for the 10-fold cross-validation on four SARS-CoV-2 RBD-ACE2 DMS datasets with  AF3 structures against experimental  structures. Higher PCC values indicate better results. (b): Comparison of  AF3-assisted MT-TopLap's    root mean square error (RMSE) for the 10-fold cross-validation on four SARS-CoV-2 RBD-ACE2 DMS datasets with AF3 structures against experimental  structures. Lower RMSE values indicate better results.
	(c): The AF3 structural predicted HK.3 RBD-ACE2 complex and its predicted DMS data (top) by AF3-assisted MT-TopLap. DMS predictions are obtained by performing a 10-fold cross-validation using the experimental DMS data (bottom). The $x$-axis labels represent the RBD residues and its wild amino acid types. The $y$-axis labels represents the mutant amino acid types.}
	\label{fig:results}
\end{figure} 

The four experimental  DMS datasets comprise of two   datasets from the original SARS-CoV-2 RBD-ACE2 DMSs  (6M0J-RBD-1, 6M0J-RBD-2) \cite{starr2020deep,linsky2020novo}, and   datasets from the BA.1-ACE2 complex (7T9L-RBD) and BA.2-ACE2 complex (7XB0-RBD) \cite{starr2022deep}. Most of the experimental DMS datasets were released during the COVID-19 pandemic to speed up the understanding of how RBD mutations affect SARS-CoV-2 infectivity and antibody resistance. One of the first SARS-CoV-2 experimental datasets used a yeast-surface-display platform to measure the expression of folded RBD protein and its binding to ACE2\cite{starr2020deep}. The per-barcode counts from these experiments were used to create a functional score, in the form of log enrichment ratios to estimate the RBD-ACE2 binding affinity values\cite{starr2020deep}. Consequently, DMS became a crucial tool for examining the SARS-CoV-2 RBD-ACE2 interaction and for designing vaccines and antibodies. Further details can be found in the Supplementary Information. 

 Figures \ref{fig:results}(a) and (b) show  the comparative analysis of the performance of AF3-assisted MT-TopLap for the four SARS-CoV-2 RBD-ACE2 DMS datasets. AF3-assisted MT-TopLap only displayed an average decrease of 1.1\% in PCC when compared to using Protein Data Bank (PDB) structures. 
Nonetheless, the four SARS-CoV-2 RBD-ACE2 datasets displayed an average of 9.3\% increase in RMSE when validated with AF3-assisted MT-TopLap. AF3-assisted MT-TopLap also achieved the best performance in the validation test for 7XB0-RBD with a PCC of 0.968, only 0.6\% lower than the performance using PDB complexes. AF3-assisted MT-TopLap's predicted BFE changes for 7XB0-RBD obtained the lowest RMSE of 0.338 kcal/mol as compared to the other three DMS datasets. This indicates that the AF3-assisted MT-TopLap is still highly effective in predicting BFE changes caused by BA.2 RBD DMS mutations.

To enhance AF3-assisted MT-TopLap's capabilities using experimental DMS data, we use transfer learning to update AF3-assisted MT-TopLap by training it with latest experimental DMS data. This process is particularly important for DMS in SARS-CoV-2 pandemic research, as experimental DMS approaches can be slow and inefficient for tracking the rapid  viral evolution. Recently, transfer learning has been applied to fine-tune MT-TopLap to predict DMS in SARS-CoV-2 S protein RBD binding to ACE2 in various animal species like bats, cats, deer, and hamsters due to RBD mutations\cite{wee2024preventing}. This enabled MT-TopLap to identify potential RBD mutations that enhances human-animal cross-transmission \cite{wee2024preventing}. With AF3, we can finetune MT-TopLap with new experimental DMS data even though high quality 3D SARS-CoV-2 RBD-ACE2 variants are unavailable in PDB. To finetune AF3-assisted MT-TopLap, we collected a recent experimental DMS dataset based on the SARS-CoV-2 HK.3 variant RBD-ACE2 complex\cite{taylor2024deep}, which does not have an available high quality 3D HK.3 RBD-human ACE2 complex in PDB (accessed PDB on 5th Oct 2024). HK.3 is one of the XBB subvariants and is known for its ``FLip" substitutions with mutations S:L455F and S:F456L. Similarly, we use the AlphaFold Server to predict the 3D structure of HK.3 variant RBD-ACE2 complex. Figure \ref{fig:results}(c) shows the RBD-ACE2 complex predicted by AF3, the comparison between experimental BFE changes (converted from enrichment ratios) (Figure \ref{fig:results}(c) bottom), and predicted BFE changes induced by HK.3 RBD mutations (Figure \ref{fig:results}(c) top). 
The predicted BFE changes are obtained by performing a 10-fold cross-validation after MT-TopLap has been pre-trained with the SARS-CoV-2 RBD-ACE2 datasets and the S8338 dataset. The patterns of the predicted DMS is observed to resemble the patterns of the converted BFE changes with a PCC of 0.81 and an RMSE of 1.10. On the whole, MT-TopLap's prediction still reflects the overall trend and this supports the   use of latest experimental DMS data and AF3 structures to update the MT-TopLap model. 

\begin{figure}[!htpb]
	\centering
	\includegraphics[width=\textwidth]{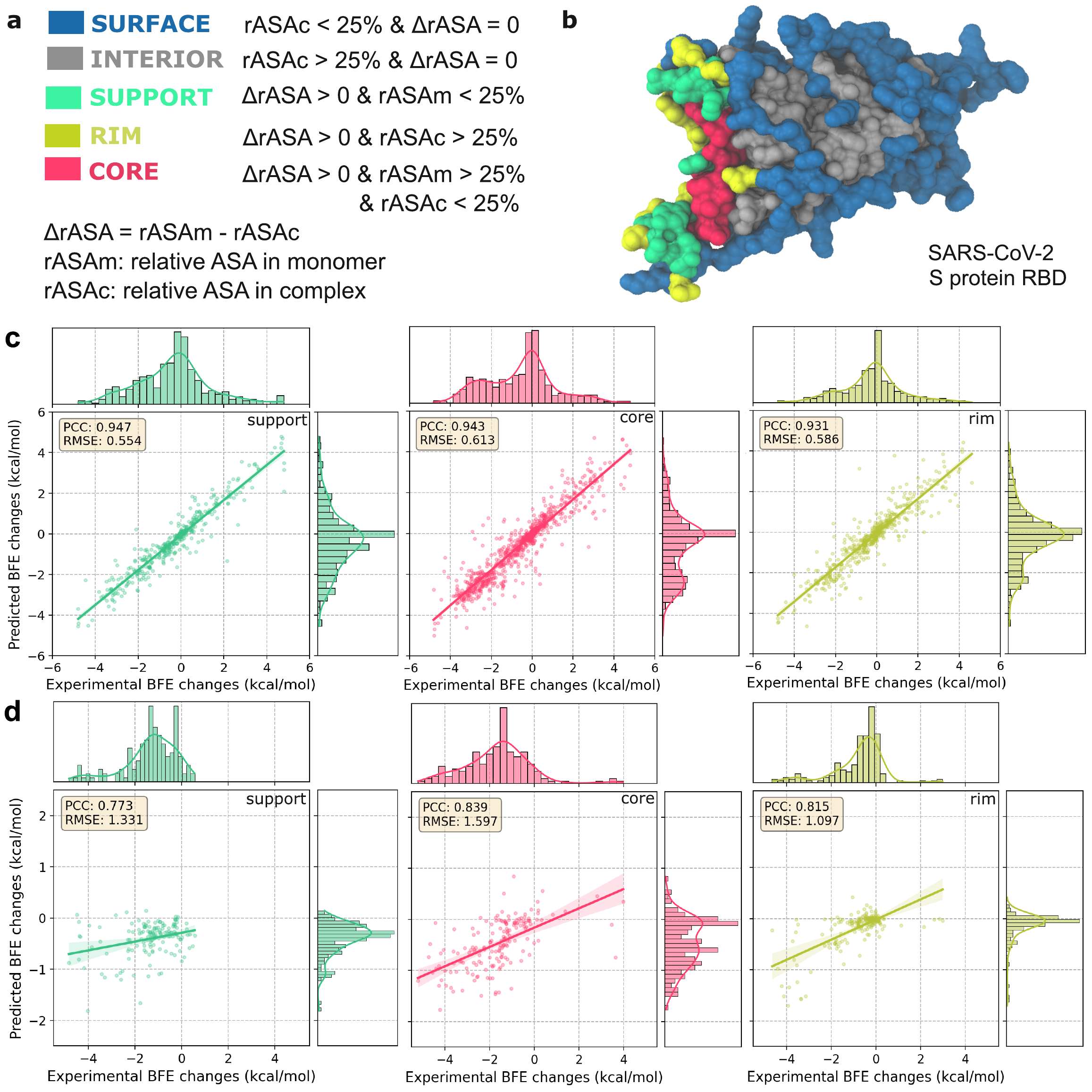}
	\caption{(a): Definitions of structural regions based on relative ASA in monomer and complex \cite{levy2010simple}.
	(b): Structural regions on the Spike protein RBD (PDBID: 6M0J \cite{lan2020structure}). Amino acids are assigned to the surface, interior, support, rim, and core based on the rASA in monomer and complex. Structures are plotted by VMD \cite{VMD}.
	(c): The 10-fold cross-validations for the four AF3-assisted SARS-CoV-2 RBD-ACE2 datasets displayed an average PCC of 0.933 and the average RMSE of 0.570 (see Figure \ref{fig:results}(a)-(b)). The combined prediction results for different residue region types for four AF3-assisted SARS-CoV-2 RBD-ACE2 datasets according to Fig. \ref{fig:results}(a) with PCCs of 0.947, 0.943, and 0.931 for the binding interfaces: support, core, and rim respectively. The average enrichment of the experimental DMS data is compared.
	(d): The 10-fold cross-validations of deep mutational scanning on the HK.3 RBD-ACE2\cite{taylor2024deep} shows a PCC of 0.81 and a RMSE of 1.10. Prediction results for different residue region types according to Fig. \ref{fig:results}(a) with PCCs of 0.773, 0.839, and 0.815 for the binding interfaces: support, core, and rim, respectively. The average enrichment of the experimental DMS data is compared.}
	\label{fig:bind_interface}
\end{figure}

\section{Discussion}

\subsection{Analysis of BFE changes by structural regions}

In this section, we analyze the results by categorizing the mutation residues based on their structural regions. The relative accessible surface area (rASA) values are calculated based on the definitions in Figure \ref{fig:bind_interface}(a) and this categorizes each mutation residue into interior and surface structural regions depending on the rASA values monomeric and PPI states\cite{levy2010simple}. Mutated residues within the binding interface significantly affect the BFE changes of the PPI and are further categorized into support, core and rim regions\cite{bogan1998anatomy}. Figure \ref{fig:bind_interface}(b) illustrates the structural regions for the original SARS-CoV-2 S protein RBD. This approach has been strongly validated in previous research as effective for identifying protein interaction interfaces. The variability in rASA calculations enables the dynamic classification of mutation residues in PPIs, even underscoring the viral adaptability in SARS-CoV-2 RBD-ACE2 interactions. 

Predicting BFE changes induced by DMS mutations within the binding interface is most vital in assessing AF3-assisted MT-TopLap's performance. In this study, experimental log enrichment ratios were converted into binding free energies, albeit with errors. Some discrepancies were noted in mutations occurring both inside and on the surface. Figure \ref{fig:bind_interface}(c)-(d) illustrates the 10-fold cross-validation performance of AF3-assisted MT-TopLap in the support, core, and rim regions of the binding interface. In Figures \ref{fig:bind_interface}(c) and (d), both the four SARS-CoV-2 RBD-ACE2 data and the HK.3 RBD-ACE2 DMS data displayed high correlations in the support, rim, and core regions. This indicates that AF3-assisted MT-TopLap performs well in predicting BFE changes upon mutations in binding interface of PPIs. The high correlations for the HK.3 RBD-ACE2 DMS data also suggest that finetuning MT-TopLap remains robust in predicting the binding interface of HK.3 RBD-ACE2 complex. 

Next, we examine the results of AF3-assisted MT-TopLap for mutated residues beyond the binding interface. Figure S2 highlights the average PCC of AF3-assisted MT-TopLap for predicted BFE changes due to surface and interior mutations. Across the four DMS datasets, high correlations were observed for surface and interior mutations (see Figure S2). However, the HK.3 RBD-ACE2 DMS dataset showed weaker PCCs, likely due to significant negative BFE changes (ranging from -5 to 0) for surface and interior mutations, while predicted values ranged from -2 to 0. Thus, the PCCs for surface and interior regions were down to 0.557 and 0.837 respectively. This finding aligns with earlier studies where the TDL-DMS model also reported weaker correlations for surface and interior mutations\cite{chen2023topological}. Nonetheless, AF3-assisted MT-TopLap performed well on the binding interface of the protein-protein complex, which is most important for understanding mutational impacts on PPI systems (see Figure \ref{fig:bind_interface}).

\subsection{Analysis of BFE changes by mutation types}

The prediction results are also analyzed by categorizing over different mutation types. The pattern  of predicted BFE changes based on various mutation types is a crucial element in protein design, particularly in the development of monoclonal antibodies (mAbs). Here, we assess how well our  predictions resemble the distribution in experimental data by examining the behavior of our model for 20 distinct amino acid types across the four SARS-CoV-2 RBD-ACE2 DMS datasets. 

\begin{figure}[!htpb]
	\centering
	\includegraphics[width=.85\textwidth]{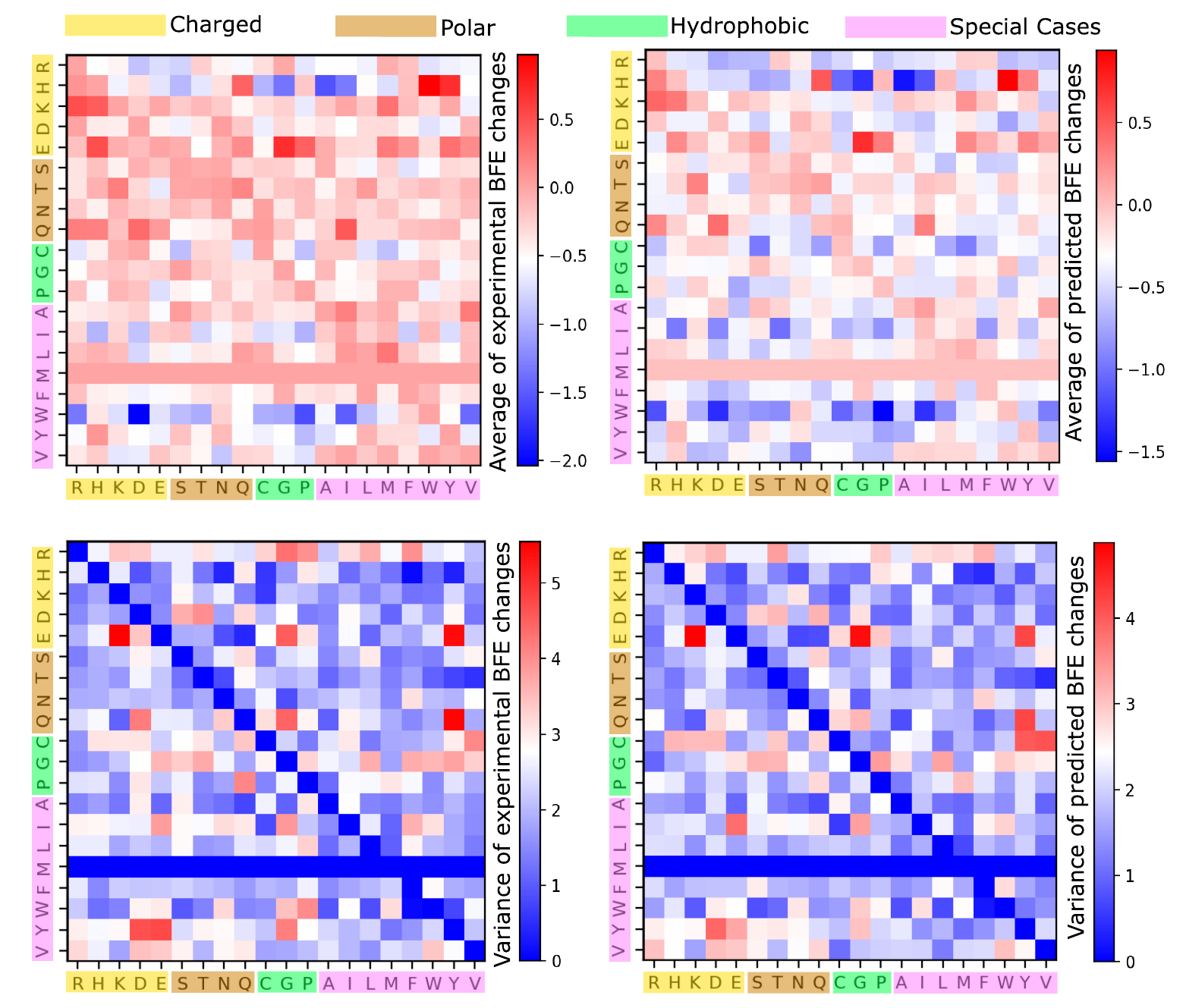}
	\caption{A comparison of average experimental and predicted BFE changes following mutations associated with different amino acid types for all the four SARS-CoV-2 RBD-ACE2 DMS datasets. The $x$-axis labels the residue type of the original RBD amino acids, whereas the $y$-axis labels the residue type of the mutant. Note that there is no amino acid MET (M) on RBD mutations. Top: Average BFE changes following mutation. Bottom: Variance of BFE changes following mutation. Left: Experimental values. Right: Predicted values.}
	\label{fig:mut_grps}
\end{figure}

Using AF3 structures, the prediction patterns remain closely aligned with the experimental data, both in terms of average BFE changes upon mutation and their variance (see Figure \ref{fig:mut_grps}). The average of predicted BFE changes upon mutation exhibits primarily a negative change, as evidenced by the predominance of negative range in the color bars. Additionally, the variance of the predicted BFE changes are generally lower than the variance of the experimental values, as indicated by the shift in the color bar range. This implies that while achieving highly accurate BFE changes upon mutation is crucial, maintaining a level of diversity comparable to the experimental data remains a challenging endeavor.

In terms of amino acid sizes, we categorize the 20 amino acids into  charged, polar, hydrophobic, and special-case groups. In Figure \ref{fig:mut_grps}, we observe that mutations from charged or polar residues to other types generally lead to the most positive BFE changes. This indicates increased stability within the PPI systems, such as when mutating from K, E and T to other amino acid types. On the other hand, higher instability is predicted for mutations from W to other residues as shown by the deep blue squares. This analysis provides insights into the molecular dynamics and potential mutation effects on SARS-CoV-2 RBD-ACE2 binding interface. 

\subsection{Tracking SARS-CoV-2 variants} 

\begin{figure}[!htpb]
	\centering
	\includegraphics[width=\textwidth]{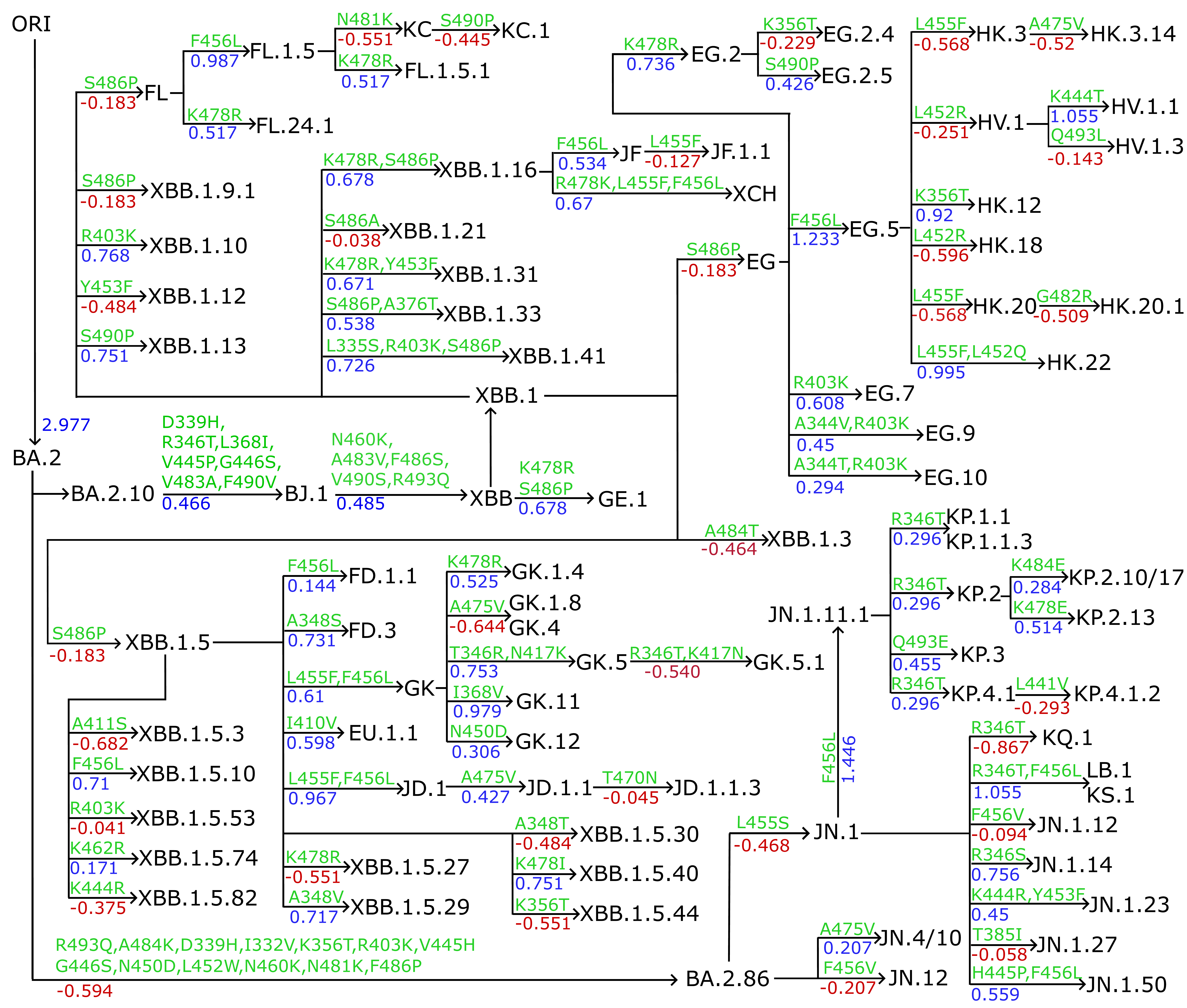}
	\caption{The lineage of Omicron XBB and BA.2.86 subvariants. RBD mutation-induced BFE changes are predicted with MT-TopLap after finetuning with HK.3 variant DMS dataset. RBD mutation-induced BFE changes (kcal/mol) are marked from parent generations to children as well as mutations. ORI represents the original SARS-CoV-2 virus.}
	\label{fig:lineage}
\end{figure}

SARS-Cov-2 evolution is driven by two natural selection mechanisms, namely infectivity strengthening \cite{chen2020mutations} and antibody resistance (or vaccine breakthrough) \cite{wang2021mechanisms}. These mechanisms were used to predict emerging dominant SARS-CoV-2 variants \cite{chen2022omicron,chen2022persistent}.  

In the results, we finetuned AF3-assisted MT-TopLap with the HK.3 variant DMS data. This expanded AF3-assisted MT-TopLap's training data but also enhanced its capabilities in tracking the recent evolutionary trajectories of SARS-CoV-2 virus. Figure \ref{fig:lineage} shows the annotation tree plots of BA.2.10 and BA.2.86 subvariants. RBD mutations from parent generations to their children are marked on the arrows and predicted binding free energy (BFE) changes (kcal/mol) induced by the corresponding RBD mutations. BFE changes from parent generations to their children are summed up by the BFE changes upon each RBD mutation. Positive BFE changes upon mutation are colored blue while negative values are colored red.

The BA.2.10 brought about the XBB wave, a series of SARS-CoV-2 variants that emerged from a recombination event between two Omicron subvariants, BA.2.10.1 and BA.2.75. The XBB lineage was first reported in India during the summer of 2022. One of the most notable variants within this lineage is XBB.1.5, also known as the ``Kraken" subvariant. It was first detected in the United States in October 2022. XBB.1.5 is highly transmissible and has shown significant immune evasion capabilities, making it a concern for global health authorities. As a result of RBD mutations, XBB.1.5 generated 24 subvariants (see Figure \ref{fig:lineage}). Thereafter, the subvariant EG.5, a descendant of XBB.1.9.2, was detected in early February 2023. EG.5 became a pre-dominant subvariant in August 2023\cite{taylor2024deep}. Notably, HK.3 is a descendant of EG.5.1 and is more transmissible due to the mutations S:L455F and S:F456L. 

Apart from the competing strains of BA.2.10, i.e., EG.5 and HK.3, a descendant of BA.2 known as BA.2.86, was detected in August 2023 and exhibits as many as 30 mutations. In Figure \ref{fig:lineage}, BA.2 requires 13 RBD mutations to become BA.2.86. It was only until early 2024 that a BA.2.86 subvariant, JN.1, became a widely circulating variant in the United States. As of 28th June 2024, both BA.2.86 and JN.1 are variants of interests (VOIs) under the WHO. As of 24th September 2024, JN.1 subvariants such as KP.2, KP.3 and LB.1 are circulating variants monitored by the WHO. 


Interestingly, the Q493E mutation has been observed to decrease BFE changes in all previous SARS-CoV-2 variants\cite{taylor2024deep}. However, in the current KP.3 variant, when Q493E is combined with L455S and F456L mutations, the effect has reversed thus promoting an increase in BFE changes\cite{taylor2024deep}. This is consistent with Figure \ref{fig:lineage} where a positive BFE change of 0.455 kcal/mol due to Q493E mutation led to the KP.3 variant. However, Figure \ref{fig:lineage} also shows that there are other subvariants of KP.2 (e.g. KP.2.10/17 and KP.2.13) which generated a higher BFE change but are currently not monitored by the WHO. Likewise, LB.1 has a lower BFE change as compared to JN.1.11.1, and yet it is a circulating variant monitored by the WHO. A recent study indicated that an epistatic drift exist in the RBD mutational effects of the current SARS-CoV-2 virus evolution \cite{taylor2024deep}. This suggests that the mutational effects on KP.2, KP.3 and LB.1 are possibly influenced by new the antibody resistance mechanism\cite{wang2021mechanisms}, which is beyond the scope of the  present work. 

\section{Methods}

Apart from the four experimental SARS-CoV-2 RBD-ACE2 DMS datasets used to validate AF3-assisted MT-TopLap, the largest mutation-induced BFE changes database, SKEMPI 2.0\cite{jankauskaite2019skempi} is also used as multi-task training data. 

As mentioned in the results section, AlphaFold Server was used to predict the 3D protein-protein complexes for the four experimental SARS-CoV-2 RBD-ACE2 DMS data. Experimental SARS-CoV-2 RBD-ACE2 DMS datasets used are SARS-CoV-2 DMS obtained from RBD-induced mutations for original SARS-CoV-2 RBD binding to ACE2\cite{starr2020deep, linsky2020novo}. Additionally, mutational scanning of BA.1 and BA.2 RBD binding to ACE2 is also included\cite{starr2022deep}.  Comprehensive details about the experimental SARS-CoV-2 RBD-ACE2 DMS datasets and the SKEMPI 2.0 database are available in the Supporting Information. 

\subsection{AF3-assisted MT-TopLap for predicting BFE changes upon mutation}

The multi-task topological Laplacian (MT-TopLap) model was built by integrating persistent Laplacian, auxiliary and pre-trained transformer features to predict PPI BFE changes upon mutation\cite{wee2024preventing}. In this work, an AF3-assisted MT-TopLap is developed to predict BFE changes upon mutation with AF3 generated PPI complex structures. As shown in Figure S1, the AF3 generated 3D SARS-CoV-2 RBD-ACE2 complex is used to generate persistent Laplacian features and auxiliary features. Persistent Laplacian features provide a topological representation, using element- and site-specific atom sets to simplify the structural complexity of protein–protein complexes and encode vital biological information into topological invariants. Previously, a major success for applying persistent Laplacian features is the ability to use topological deep learning model to accurately predict the dominance of Omicron BA.4 and BA.5 nearly two months before WHO made the official announcement\cite{chen2022persistent}. The ESM-2 pre-trained transformer is used to convert amino acid sequence of the PPI into a pre-trained transformer embedding. All these features are concatenated and fed into MT-TopLap to predict the BFE change upon mutation.

The deep neural network model in MT-TopLap's architecture consists of six hidden layers with 15,000 neurons in each layer and generates an output channel for each dataset. For the validation process, we perform dataset-level 10-fold cross-validation on four experimental SARS-CoV-2 RBD-ACE2 datasets. The learning rate is set to 0.0001 and 500 epochs are used for the pre-training while 200 epochs are used for the validation step. The output for each dataset is represented by an individual output channel for training the experimental enrichment ratios (see Figure S1). 
For the fine-tuning process, the weights and biases for the 2nd and third hidden layer are froze. Here, the learning rate is set to 0.0001 and 200 epochs are used for fine-tuning. Details of the method can be found in the literature \cite{wee2024preventing}. Specific details for 10-fold cross-validations and the finetuning performed in this work are given in the Supporting Information. 

\subsection{Persistent Laplacian Features}

One of the main descriptors used in AF3-assisted MT-TopLap is the persistent Laplacian feature vectors. We now outline the key mathematical principles behind the persistent Laplacian descriptors in AF3-assisted MT-TopLap. This involves describing the simplicial complex and persistent Laplacian methods, emphasizing their importance in capturing harmonic and non-harmonic spectral properties crucial for characterizing SARS-CoV-2 RBD-ACE2 interactions. Details about the auxiliary, persistent homology and the ESM-2 transformer features can be found in the Supporting Information.

To construct persistent Laplacian features for the RBD-ACE2 complex, we categorize the atoms in the complex into specific site-based subsets. These include mutation site atoms $\mathcal{A}_m$, atoms within a distance $r$ of mutation site $\mathcal{A}_{mn}(r)$, atoms from the RBD binding site $\mathcal{A}_{\text{RBD}}(r)$, and atoms from the ACE2 binding site $\mathcal{A}_{\text{ACE2}}(r)$. We also classify atoms into different element-specific subsets, such as $\{\text{C, N, O}\}$, $\mathcal{A}_{\text{ele}}$. These partitions are key for PPI model characterization, as different atom combinations capture various interaction types. For instance, subsets $\mathcal{A}_{\text{C}}\cap \mathcal{A}_{\text{RBD}}(r)$ and $\mathcal{A}_{\text{C}}\cap \mathcal{A}_{\text{ACE2}}(r)$ form hydrophobic C-C RBD-ACE2 interactions, while $\mathcal{A}_{\text{N}}\cap \mathcal{A}_{\text{RBD}}(r)$ and $\mathcal{A}_{\text{O}}\cap \mathcal{A}_{\text{ACE2}}(r)$ result in hydrophilic N-O RBD-ACE2 interactions.

In order to characterize the RBD-ACE2 binding interactions, we also modify the standard Euclidean distance matrix ${\rm DE}$ such that it excludes interactions between both atoms found in the RBD or both in the ACE2. Specifically, for interactions between atoms $A_i$ and $A_j$ in sets $\mathcal{A}$ and $\mathcal{B}$, ${\rm DI}$ is defined as follows:
\begin{equation}
{\rm DI}(A_i, A_j) = \begin{cases}
\infty, & \text{ if }A_i,A_j \in \mathcal{A} \text{ or } \text{ if }A_i,A_j \in \mathcal{B}\\
{\rm DE}(A_i, A_j), & \text{otherwise}.
\end{cases}
\end{equation}
where ${\rm DE}(\cdot, \cdot)$ is the Euclidean distance between the two atoms.

For each site and element-specific subset, the 3D atom positions generate point clouds which are then used to build simplicial complexes. A set of $k+1$ atoms from a site/element-specific subset forms $k+1$ independent points, denoted as $S=\{v_0, v_1, v_2, \cdots, v_k\}$. The convex hull of $k+1$ affinely independent points forms a $k$-simplex: with a point being a 0-simplex, an edge being a 1-simplex, a triangle being a 2-simplex, and a tetrahedron being a 3-simplex, while higher dimensions form k-simplices. A simplicial complex is created from the aggregation of these finite simplices\cite{munkres2018elements,zomorodian2005topology,Edelsbrunner:2010,Mischaikow:2013}. There are various methods for constructing simplicial complexes. For generating our persistent Laplacian-based features, we utilized the Vietoris-Rips complex for dimension 0 and the Alpha complex for dimensions 1 and 2. The Vietoris-Rips (VR) complex forms simplices by connecting subsets of points with diameters not exceeding a given threshold. On the other hand, the Alpha complex is derived from Delaunay triangulation, constrained by a radius not exceeding a specified threshold, which subdivides the convex hull of a point set into triangles.

For a simplicial complex $K$, a $k$-th chain $c_k$ is the formal sum of $k$-simplicies in $K$, i.e. $c_k = \sum_i \alpha_i\sigma_i^k$. A boundary operator $\partial_k: C_k \rightarrow C_{k-1}$ defined on a $k$-th chain $c_k$ is
\begin{equation*}
\partial_k c_k = \sum_{i=0}^k \alpha_i\partial_{k}\sigma_i^k,
\end{equation*}
such that the boundary of a boundary is empty, i.e. $\partial_{k-1}\partial_k = \varnothing $. By defining the adjoint of $\partial_k$, i.e. $\partial_{k}^*:C_{k-1}\to C_{k}$, we have $\partial_{k}^*$ satisfying the inner product relation $\langle \partial_k(f), g \rangle=\langle f, \partial_{k}^*(g) \rangle,$ for every $f \in C_{k}$, $g \in C_{k-1}$. Then the $k$-combinatorial Laplacian or the topological Laplacian is a linear operator $\Delta_k: C_k(K) \rightarrow C_k(K)$
\begin{equation}
\Delta_k := \partial_{k+1}\partial_{k+1}^* + \partial_{k}^*\partial_{k}.
\end{equation}

In terms of matrix representations, we define $\mathbf{B}_k$ to be an $m\times n$ matrix representation of the boundary operator under the standard bases $\{\sigma_i^k\}^n_{i=1}$ and $\{\sigma_j^{k-1}\}^m_{j=1}$ of $C_k$ and $C_{k-1}$. Similarly, the matrix representation of $\partial_{k}^*$ is the transpose matrix $\mathbf{B}_k^\top$, with respect to the same ordered bases of the boundary operator $\partial_k$. Hence, the $k$-combinatorial Laplacian exhibits an $n\times n$ matrix representation $\mathbf{L}_k$ and is given by 
\begin{equation}
\mathbf{L}_k = \mathbf{B}_{k+1}\mathbf{B}_{k+1}^\top + \mathbf{B}_k^\top \mathbf{B}_k. 
\end{equation}
In the case $k=0$, then $\mathbf{L}_0 = \mathbf{B}_1\mathbf{B}_1^\top$ since $\partial_{0}$ is a zero map.

In our model, the eigenvalues of combinatorial Laplacian matrices are key topological features, independent of orientation choice\cite{horak2013spectra}. The multiplicity of zero eigenvalues of $\mathbf{L}_k$ corresponds to the $k$-th Betti number $\beta_k$, which describes the $k$-dimensional holes in a simplicial complex \cite{eckmann1944harmonische}. Specifically, $\beta_0$, $\beta_1$, and $\beta_2$ represent the number of independent components, loops, and voids, respectively. These Betti numbers provide critical insights into the fundamental structure of the protein-protein interaction, identifying loops and voids within the PPI system.

Zero eigenvalues in the Laplacian matrix signify the harmonic spectra, representing stable features like connected components and cycles that persist across different scales. Non-zero eigenvalues represent the non-harmonic spectra, revealing more transient and intricate details of the molecular shape and interactions that are not captured by Betti numbers alone. These non-harmonic spectra offer additional homotopic shape information, crucial for a comprehensive understanding of the biomolecular interaction dynamics. 

One simplicial complex is insufficient to capture all topological information from a protein-protein interaction structure. By integrating combinatorial Laplacian and multiscale filtration, we track changes in harmonic and non-harmonic spectra by varying a filtration parameter such as radii/diameter for the VR complex. For an oriented simplicial complex $K$, filtration generates a nested sequence of simplicial complexes $(K_t)^m_{t=0}$: $$ \varnothing = K_0 \subseteq K_1 \subseteq \cdots \subseteq K_m=K. $$ Persistent Laplacian (PL) produces a sequence of simplicial complexes as the filtration parameter increases, allowing us to generate a sequence of combinatorial Laplacian matrices $\mathbf{L}_{k}^0,\mathbf{L}_{k}1,\mathbf{L}_{k}2,\mathbf{L}_{k}^3,\cdots,\mathbf{L}_{k}^n$, where $\mathbf{L}_{k}^t=\mathbf{L}_k(K_t)$. By altering the filtration parameter and performing diagonalization on the $k$-combinatorial Laplacian matrix, we can examine the topology and spectrum characteristics. The eigenvalues of $\mathbf{L}_k(K_t)$ can be arranged in ascending order: $$ \text{Spectra}(\mathbf{L}_k^t) = \{(\lambda_1)_k^t, (\lambda_2)_k^t, \cdots, (\lambda_n)_k^t\}, $$ where $\mathbf{L}_k^t$ is an $n\times n$ matrix. Additionally, the $p$-persistent $k$-combinatorial Laplacian can be extended based on the boundary operator.

When generating features, we take both harmonic and non-harmonic spectra into account for each persistent Laplacian in zero dimensions. Using filtration with the Rips complex and DI distance, we generate 0-dimensional PL features ranging from 0\AA\space to 6\AA\space with a grid size of 0.5\AA. For non-harmonic spectra, we count the occurrences and compute seven statistical values: sum, minimum, maximum, mean, standard deviation, variance, and the sum of squared eigenvalues. This results in eight statistical values for each of the nine atomic pairs, producing a total of 72 features for a protein in zero dimensions. When concatenated for different dimensions of the wild type and mutant, the total 0-dimensional PL-based feature size is 1872.

For one- and two-dimensional PL features, we use the Alpha complex with DE distance for filtration. Due to the limited number of atoms in local protein structures that can form only a few high-dimensional simplexes, we focus on the harmonic spectra of persistent Laplacians. This captures topological invariants of high-dimensional interactions, providing a comprehensive view of the molecular structure. Using GUDHI\cite{maria2014gudhi}, the persistence of harmonic spectra is represented by persistent barcodes. We generate topological feature vectors by computing statistics of bar lengths, births, and deaths, excluding bars shorter than 0.1\AA\space as they lack clear physical significance. The statistics computed include the sum, maximum, and mean of bar lengths; minimum and maximum of bar birth values; and minimum and maximum of bar death values. Each set of point clouds results in a seven-dimensional vector. These features are calculated for nine single atomic pairs and one heavy atom pair, yielding 140 features for one- and two-dimensional PL vectors of a protein. When combined for different dimensions of wild type, mutant, and their differences, the total higher-dimensional PL-based feature size is 420.

The persistent Laplacian features generated in this manner provide a rich and detailed representation of the RBD-ACE2 interactions, capturing both local and global structural features that are critical for accurate modeling. These features enable us to perform critical analyses based on the predicted BFE changes caused by RBD mutations, thereby leveraging the deep contextual information encoded in the RBD-ACE2 binding domain. By integrating harmonic and non-harmonic spectra and considering different dimensions, we obtain a comprehensive understanding of the topological and geometric characteristics of the SARS-CoV-2 RBD-ACE2 system, facilitating accurate predictions and insights into its behavior and interactions. Further details about persistent Laplacian methods and their applications are reported in \cite{wang2020persistent}.

\section{Conclusion}

Our ability to rapidly respond to viral evolution underpins the health and well-being of the human race. A topological deep learning (TDL)-based approach has a proven track record in forecasting emerging dominant SARS-CoV-2 variants, such as Omicron BA.2 \cite{chen2022omicron} and BA.4/BA.5 \cite{chen2022persistent}, approximately two months in advance. However, such forecasting is often bottlenecked by the lack of experimental deep mutational scanning (DMS) data and the three-dimensional (3D) structures of protein-protein interaction (PPI) complexes for fast evolving viruses. We address this challenge by leveraging AlphaFold 3 (AF3)  to predict the  PPI complexes. Specifically, we propose an AF3-assisted multi-task topological Laplacian (MT-TopLap) model to predict DMS data and forecast emerging viral variants. AF3-assisted MT-TopLap leverages TDL and persistent Laplacians (PL) to capture both topological and geometric features of PPI complexes. It  accurately predicts both the DMS profiles and the binding free energy (BFE) changes of the PPI complexes for viral variants. We validate the proposed AF3-assisted MT-TopLap strategy by using four experimental DMS datasets on the SARS-CoV-2 spike receptor binding domain (RBD) bound to the human angiotensin-converting enzyme 2 (ACE2). 
With 3D PPI complexes generated by AF3, AF3-assisted MT-TopLap achieves robust predictive performance, showing minimal deviation from results obtained using high-quality experimental 3D structures. Our model's success extends to new variants, including the SARS-CoV-2 HK.3 variant, where it achieved a Pearson correlation coefficient (PCC) of 0.81. This highlights AF3-assisted MT-TopLap's ability to adapt to emerging viral data and respond effectively to fast viral evolution.


%
%

\section*{Data Availability}
The training data used in this work consists of the comprehensive four SARS-CoV-2 RBD-ACE2 datasets which are readily available in \linebreak  \href{https://github.com/ExpectozJJ/MT-TopLap/tree/main/AF3\_SARS-CoV-2}{https://github.com/ExpectozJJ/MT-TopLap/tree/main/AF3\_SARS-CoV-2}.
The AF3 SARS-CoV-2 RBD-ACE2 complexes can also be downloaded from \linebreak \href{https://github.com/ExpectozJJ/MT-TopLap/tree/main/AF3\_SARS-CoV-2}{https://github.com/ExpectozJJ/MT-TopLap/tree/main/AF3\_SARS-CoV-2}. The original PDB files used in this work can be downloaded from the official Protein Databank: \linebreak \href{https://www.rcsb.org/}{https://www.rcsb.org/}. The SKEMPI 2.0 database is also readily available from \linebreak  \href{https://life.bsc.es/pid/skempi2}{https://life.bsc.es/pid/skempi2}. 

\section*{Code Availability}
The codes for 10-fold cross-validation and finetuning tasks can be obtained from the following source: \href{https://github.com/ExpectozJJ/MT-TopLap/tree/main/AF3\_SARS-CoV-2}{https://github.com/ExpectozJJ/MT-TopLap/tree/main/AF3\_SARS-CoV-2}. Details about the feature generation and MT-TopLap's model architecture is available in Supporting Information.

\section*{Acknowledgments}
This work was supported in part by NIH grants  R01GM126189, R01AI164266, and R35GM148196, NSF grants DMS-2052983   and IIS-1900473,    MSU Research Foundation,  and Bristol-Myers Squibb 65109.

\vspace{0.6cm}
\bibliographystyle{ieeetr}
\bibliography{refs}

\begin{thebibliography}{10}

\bibitem{flu-vaccines-work2024}
CDC, ``{CDC Seasonal Flu Vaccine Effectiveness Studies},'' 2024.
\newblock
  \url{https://www.cdc.gov/flu-vaccines-work/php/effectiveness-studies/index.html}
  [Accessed: (November 15, 2024)].

\bibitem{chen2022omicron}
J.~Chen and G.-W. Wei, ``Omicron {BA. 2 (B. 1.1. 529.2)}: high potential for
  becoming the next dominant variant,'' {\em The journal of physical chemistry
  letters}, vol.~13, no.~17, pp.~3840--3849, 2022.

\bibitem{chen2022persistent}
J.~Chen, Y.~Qiu, R.~Wang, and G.-W. Wei, ``Persistent {Laplacian projected
  Omicron BA. 4 and BA. 5} to become new dominating variants,'' {\em Computers
  in Biology and Medicine}, vol.~151, p.~106262, 2022.

\bibitem{chen2020mutations}
J.~Chen, R.~Wang, M.~Wang, and G.-W. Wei, ``Mutations strengthened {SARS-CoV}-2
  infectivity,'' {\em Journal of molecular biology}, vol.~432, no.~19,
  pp.~5212--5226, 2020.

\bibitem{wang2021mechanisms}
R.~Wang, J.~Chen, and G.-W. Wei, ``Mechanisms of sars-cov-2 evolution revealing
  vaccine-resistant mutations in europe and america,'' {\em The journal of
  physical chemistry letters}, vol.~12, no.~49, pp.~11850--11857, 2021.

\bibitem{starr2022deep}
T.~N. Starr, A.~J. Greaney, C.~M. Stewart, A.~C. Walls, W.~W. Hannon,
  D.~Veesler, and J.~D. Bloom, ``Deep mutational scans for {ACE2} binding,
  {RBD} expression, and antibody escape in the {SARS-CoV-2 Omicron BA. 1 and
  BA. 2} receptor-binding domains,'' {\em PLoS pathogens}, vol.~18, no.~11,
  p.~e1010951, 2022.

\bibitem{starr2020deep}
T.~N. Starr, A.~J. Greaney, S.~K. Hilton, D.~Ellis, K.~H. Crawford, A.~S.
  Dingens, M.~J. Navarro, J.~E. Bowen, M.~A. Tortorici, A.~C. Walls, {\em
  et~al.}, ``Deep mutational scanning of {SARS-CoV-2} receptor binding domain
  reveals constraints on folding and {ACE2} binding,'' {\em Cell}, vol.~182,
  no.~5, pp.~1295--1310, 2020.

\bibitem{dadonaite2024spike}
B.~Dadonaite, J.~Brown, T.~E. McMahon, A.~G. Farrell, M.~D. Figgins,
  D.~Asarnow, C.~Stewart, J.~Lee, J.~Logue, T.~Bedford, {\em et~al.}, ``Spike
  deep mutational scanning helps predict success of {SARS-CoV-2} clades,'' {\em
  Nature}, vol.~631, no.~8021, pp.~617--626, 2024.

\bibitem{bergquist2020covid}
S.~Bergquist, T.~Otten, and N.~Sarich, ``{COVID-19} pandemic in the united
  states,'' {\em Health policy and technology}, vol.~9, no.~4, pp.~623--638,
  2020.

\bibitem{geng2019finding}
C.~Geng, L.~C. Xue, J.~Roel-Touris, and A.~M. Bonvin, ``Finding the
  {$\Delta\Delta$G spot: Are} predictors of binding affinity changes upon
  mutations in protein--protein interactions ready for it?,'' {\em Wiley
  Interdisciplinary Reviews: Computational Molecular Science}, vol.~9, no.~5,
  p.~e1410, 2019.

\bibitem{pandurangan2020prediction}
A.~P. Pandurangan and T.~L. Blundell, ``Prediction of impacts of mutations on
  protein structure and interactions: {SDM}, a statistical approach, and
  m{CSM}, using machine learning,'' {\em Protein Science}, vol.~29, no.~1,
  pp.~247--257, 2020.

\bibitem{geng2019isee}
C.~Geng, A.~Vangone, G.~E. Folkers, L.~C. Xue, and A.~M. Bonvin, ``i{SEE}:
  Interface structure, evolution, and energy-based machine learning predictor
  of binding affinity changes upon mutations,'' {\em Proteins: Structure,
  Function, and Bioinformatics}, vol.~87, no.~2, pp.~110--119, 2019.

\bibitem{li2021saambe}
G.~Li, S.~Pahari, A.~K. Murthy, S.~Liang, R.~Fragoza, H.~Yu, and E.~Alexov,
  ``{SAAMBE-SEQ}: a sequence-based method for predicting mutation effect on
  protein--protein binding affinity,'' {\em Bioinformatics}, vol.~37, no.~7,
  pp.~992--999, 2021.

\bibitem{siebenmorgen2020computational}
T.~Siebenmorgen and M.~Zacharias, ``Computational prediction of
  protein--protein binding affinities,'' {\em Wiley Interdisciplinary Reviews:
  Computational Molecular Science}, vol.~10, no.~3, p.~e1448, 2020.

\bibitem{dehouck2013beatmusic}
Y.~Dehouck, J.~M. Kwasigroch, M.~Rooman, and D.~Gilis, ``Be{AtMuSiC}:
  prediction of changes in protein--protein binding affinity on mutations,''
  {\em Nucleic acids research}, vol.~41, no.~W1, pp.~W333--W339, 2013.

\bibitem{romero2022ppi}
S.~Romero-Molina, Y.~B. Ruiz-Blanco, J.~Mieres-Perez, M.~Harms, J.~M\"unch,
  M.~Ehrmann, and E.~Sanchez-Garcia, ``{PPI}-affinity: A web tool for the
  prediction and optimization of protein--peptide and protein--protein binding
  affinity,'' {\em Journal of proteome research}, vol.~21, no.~8,
  pp.~1829--1841, 2022.

\bibitem{xue2016prodigy}
L.~C. Xue, J.~P. Rodrigues, P.~L. Kastritis, A.~M. Bonvin, and A.~Vangone,
  ``{PRODIGY}: a web server for predicting the binding affinity of
  protein--protein complexes,'' {\em Bioinformatics}, vol.~32, no.~23,
  pp.~3676--3678, 2016.

\bibitem{yang2023area}
Y.~X. Yang, J.~Y. Huang, P.~Wang, and B.~T. Zhu, ``Area-affinity: A web server
  for machine learning-based prediction of protein--protein and
  antibody--protein antigen binding affinities,'' {\em Journal of Chemical
  Information and Modeling}, vol.~63, no.~11, pp.~3230--3237, 2023.

\bibitem{wang2020topology}
M.~Wang, Z.~Cang, and G.-W. Wei, ``A topology-based network tree for the
  prediction of protein--protein binding affinity changes following mutation,''
  {\em Nature Machine Intelligence}, vol.~2, no.~2, pp.~116--123, 2020.

\bibitem{cang2017topologynet}
Z.~Cang and G.-W. Wei, ``Topology{N}et: {T}opology based deep convolutional and
  multi-task neural networks for biomolecular property predictions,'' {\em PLoS
  computational biology}, vol.~13, no.~7, p.~e1005690, 2017.

\bibitem{edelsbrunner2008persistent}
H.~Edelsbrunner, J.~Harer, {\em et~al.}, ``Persistent homology-a survey,'' {\em
  Contemporary mathematics}, vol.~453, no.~26, pp.~257--282, 2008.

\bibitem{zomorodian2004computing}
A.~Zomorodian and G.~Carlsson, ``Computing persistent homology,'' in {\em
  Proceedings of the twentieth annual symposium on Computational geometry},
  pp.~347--356, 2004.

\bibitem{chen2021prediction}
J.~Chen, K.~Gao, R.~Wang, and G.-W. Wei, ``Prediction and mitigation of
  mutation threats to {COVID}-19 vaccines and antibody therapies,'' {\em
  Chemical science}, vol.~12, no.~20, pp.~6929--6948, 2021.

\bibitem{wang2020persistent}
R.~Wang, D.~D. Nguyen, and G.-W. Wei, ``Persistent spectral graph,'' {\em
  International journal for numerical methods in biomedical engineering},
  vol.~36, no.~9, p.~e3376, 2020.

\bibitem{wei2024persistent}
X.~Wei and G.-W. Wei, ``Persistent sheaf laplacians,'' {\em Foundations of Data
  Science}, 2024.

\bibitem{shen2024persistent}
L.~Shen, J.~Liu, and G.-W. Wei, ``Persistent mayer homology and persistent
  mayer laplacian,'' {\em Foundations of Data Science}, vol.~6, no.~4,
  pp.~584--612, 2024.

\bibitem{wee2022persistent}
J.~Wee and K.~Xia, ``Persistent spectral based ensemble learning
  ({PerSpect-EL}) for protein--protein binding affinity prediction,'' {\em
  Briefings in Bioinformatics}, vol.~23, no.~2, p.~bbac024, 2022.

\bibitem{liu2022hom}
X.~Liu, H.~Feng, J.~Wu, and K.~Xia, ``Hom-complex-based machine learning
  ({HCML}) for the prediction of protein--protein binding affinity changes upon
  mutation,'' {\em Journal of chemical information and modeling}, vol.~62,
  no.~17, pp.~3961--3969, 2022.

\bibitem{rives2021biological}
A.~Rives, J.~Meier, T.~Sercu, S.~Goyal, Z.~Lin, J.~Liu, D.~Guo, M.~Ott, C.~L.
  Zitnick, J.~Ma, {\em et~al.}, ``Biological structure and function emerge from
  scaling unsupervised learning to 250 million protein sequences,'' {\em
  Proceedings of the National Academy of Sciences}, vol.~118, no.~15,
  p.~e2016239118, 2021.

\bibitem{wee2024integration}
J.~Wee, J.~Chen, K.~Xia, and G.-W. Wei, ``Integration of persistent {L}aplacian
  and pre-trained transformer for protein solubility changes upon mutation,''
  {\em Computers in Biology and Medicine}, p.~107918, 2024.

\bibitem{jankauskaite2019skempi}
J.~Jankauskait{\.e}, B.~Jim{\'e}nez-Garc{\'\i}a, J.~Dapk{\=u}nas,
  J.~Fern{\'a}ndez-Recio, and I.~H. Moal, ``{SKEMPI} 2.0: an updated benchmark
  of changes in protein--protein binding energy, kinetics and thermodynamics
  upon mutation,'' {\em Bioinformatics}, vol.~35, no.~3, pp.~462--469, 2019.

\bibitem{wee2024preventing}
J.~Wee, J.~Chen, and G.-W. Wei, ``Preventing future zoonosis: {SARS-CoV-2}
  mutations enhance human--animal cross-transmission,'' {\em Computers in
  Biology and Medicine}, vol.~182, p.~109101, 2024.

\bibitem{rodrigues2019mcsm}
C.~H. Rodrigues, Y.~Myung, D.~E. Pires, and D.~B. Ascher, ``m{CSM-PPI2}:
  {P}redicting the effects of mutations on protein--protein interactions,''
  {\em Nucleic acids research}, vol.~47, no.~W1, pp.~W338--W344, 2019.

\bibitem{abramson2024accurate}
J.~Abramson, J.~Adler, J.~Dunger, R.~Evans, T.~Green, A.~Pritzel,
  O.~Ronneberger, L.~Willmore, A.~J. Ballard, J.~Bambrick, {\em et~al.},
  ``Accurate structure prediction of biomolecular interactions with {A}lphafold
  3,'' {\em Nature}, pp.~1--3, 2024.

\bibitem{jumper2021highly}
J.~Jumper, R.~Evans, A.~Pritzel, T.~Green, M.~Figurnov, O.~Ronneberger,
  K.~Tunyasuvunakool, R.~Bates, A.~{\v{Z}}{\'\i}dek, A.~Potapenko, {\em
  et~al.}, ``Highly accurate protein structure prediction with {AlphaF}old,''
  {\em Nature}, vol.~596, no.~7873, pp.~583--589, 2021.

\bibitem{wang2019improved}
T.~Wang, Y.~Qiao, W.~Ding, W.~Mao, Y.~Zhou, and H.~Gong, ``Improved fragment
  sampling for ab initio protein structure prediction using deep neural
  networks,'' {\em Nature Machine Intelligence}, vol.~1, no.~8, pp.~347--355,
  2019.

\bibitem{wei2019protein}
G.-W. Wei, ``Protein structure prediction beyond {AlphaF}old,'' {\em Nature
  Machine Intelligence}, vol.~1, no.~8, pp.~336--337, 2019.

\bibitem{qiu2023persistent}
Y.~Qiu and G.-W. Wei, ``Persistent spectral theory-guided protein
  engineering,'' {\em Nature Computational Science}, vol.~3, no.~2,
  pp.~149--163, 2023.

\bibitem{lin2023evolutionary}
Z.~Lin, H.~Akin, R.~Rao, B.~Hie, Z.~Zhu, W.~Lu, N.~Smetanin, R.~Verkuil,
  O.~Kabeli, Y.~Shmueli, {\em et~al.}, ``Evolutionary-scale prediction of
  atomic-level protein structure with a language model,'' {\em Science},
  vol.~379, no.~6637, pp.~1123--1130, 2023.

\bibitem{mirdita2022colabfold}
M.~Mirdita, K.~Sch{\"u}tze, Y.~Moriwaki, L.~Heo, S.~Ovchinnikov, and
  M.~Steinegger, ``Colab{F}old: making protein folding accessible to all,''
  {\em Nature methods}, vol.~19, no.~6, pp.~679--682, 2022.

\bibitem{tunyasuvunakool2021highly}
K.~Tunyasuvunakool, J.~Adler, Z.~Wu, T.~Green, M.~Zielinski,
  A.~{\v{Z}}{\'\i}dek, A.~Bridgland, A.~Cowie, C.~Meyer, A.~Laydon, {\em
  et~al.}, ``Highly accurate protein structure prediction for the human
  proteome,'' {\em Nature}, vol.~596, no.~7873, pp.~590--596, 2021.

\bibitem{dejnirattisai2022sars}
W.~Dejnirattisai, J.~Huo, D.~Zhou, J.~Zahradn{\'\i}k, P.~Supasa, C.~Liu, H.~M.
  Duyvesteyn, H.~M. Ginn, A.~J. Mentzer, A.~Tuekprakhon, {\em et~al.},
  ``S{ARS-CoV-2 Omicron-B. 1.1. 529} leads to widespread escape from
  neutralizing antibody responses,'' {\em Cell}, vol.~185, no.~3, pp.~467--484,
  2022.

\bibitem{nunes2023alphafold2}
A.~Nunes-Alves and K.~Merz, ``Alpha{Fold2 in Molecular Discovery},'' 2023.

\bibitem{taylor2024deep}
A.~L. Taylor and T.~N. Starr, ``Deep mutational scanning of {SARS-CoV-2 Omicron
  BA. 2.86 and epistatic emergence of the KP. 3 variant},'' {\em Virus
  Evolution}, vol.~10, no.~1, p.~veae067, 2024.

\bibitem{linsky2020novo}
T.~W. Linsky, R.~Vergara, N.~Codina, J.~W. Nelson, M.~J. Walker, W.~Su, C.~O.
  Barnes, T.-Y. Hsiang, K.~Esser-Nobis, K.~Yu, {\em et~al.}, ``De novo design
  of potent and resilient {hACE2} decoys to neutralize {SARS-CoV-2},'' {\em
  Science}, vol.~370, no.~6521, pp.~1208--1214, 2020.

\bibitem{levy2010simple}
E.~D. Levy, ``A simple definition of structural regions in proteins and its use
  in analyzing interface evolution,'' {\em Journal of molecular biology},
  vol.~403, no.~4, pp.~660--670, 2010.

\bibitem{lan2020structure}
J.~Lan, J.~Ge, J.~Yu, S.~Shan, H.~Zhou, S.~Fan, Q.~Zhang, X.~Shi, Q.~Wang,
  L.~Zhang, {\em et~al.}, ``Structure of the {SARS-CoV-2} spike
  receptor-binding domain bound to the {ACE2} receptor,'' {\em nature},
  vol.~581, no.~7807, pp.~215--220, 2020.

\bibitem{VMD}
W.~Humphrey, A.~Dalke, and K.~Schulten, ``{VMD} -- visual molecular dynamics,''
  {\em Journal of Molecular Graphics}, vol.~14, no.~1, pp.~33--38, 1996.

\bibitem{bogan1998anatomy}
A.~A. Bogan and K.~S. Thorn, ``Anatomy of hot spots in protein interfaces,''
  {\em Journal of molecular biology}, vol.~280, no.~1, pp.~1--9, 1998.

\bibitem{chen2023topological}
J.~Chen, D.~R. Woldring, F.~Huang, X.~Huang, and G.-W. Wei, ``Topological deep
  learning based deep mutational scanning,'' {\em Computers in Biology and
  Medicine}, vol.~164, p.~107258, 2023.

\bibitem{munkres2018elements}
J.~R. Munkres, {\em Elements of algebraic topology}.
\newblock CRC Press, 2018.

\bibitem{zomorodian2005topology}
A.~J. Zomorodian, {\em Topology for computing}, vol.~16.
\newblock Cambridge university press, 2005.

\bibitem{Edelsbrunner:2010}
H.~Edelsbrunner and J.~Harer, {\em Computational topology: an introduction}.
\newblock American Mathematical Soc., 2010.

\bibitem{Mischaikow:2013}
K.~Mischaikow and V.~Nanda, ``Morse theory for filtrations and efficient
  computation of persistent homology,'' {\em Discrete and Computational
  Geometry}, vol.~50, no.~2, pp.~330--353, 2013.

\bibitem{horak2013spectra}
D.~Horak and J.~Jost, ``Spectra of combinatorial {Laplace} operators on
  simplicial complexes,'' {\em Advances in Mathematics}, vol.~244,
  pp.~303--336, 2013.

\bibitem{eckmann1944harmonische}
B.~Eckmann, ``Harmonische funktionen und randwertaufgaben in einem komplex,''
  {\em Commentarii Mathematici Helvetici}, vol.~17, no.~1, pp.~240--255, 1944.

\bibitem{maria2014gudhi}
C.~Maria, J.-D. Boissonnat, M.~Glisse, and M.~Yvinec, ``The {GUDHI} library:
  {S}implicial complexes and persistent homology,'' in {\em Mathematical
  Software--ICMS 2014: 4th International Congress, Seoul, South Korea, August
  5-9, 2014. Proceedings 4}, pp.~167--174, Springer, 2014.

\end{thebibliography}



\end{document}